\documentclass{ws-procs975x65}
\usepackage{cleveref}
\newcommand{\mr}[1]{\mathrm{#1}}

\usepackage{amsmath,amssymb,lmodern} 
\usepackage{tabularx}
\usepackage{makecell}
\usepackage{sidecap}
\usepackage[caption=false]{subfig}

\begin{document}

\title{$p$-form quintessence: exploring dark energy of $p-$forms coupled to a scalar field}

\author{Alejandro Guarnizo$^{1, 2 \, *}$, Juan P. Beltr\'an Almeida$^{1}$ and C\'esar A. Valenzuela-Toledo$^{2}$} 

\address{$^{1}$ Departamento de F\'isica, Universidad Antonio Nari\~no, \\ 
Cra 3 Este \# 47A-15, 110231, Bogot\'a DC, Colombia\\
$^{2}$ Departamento de F\'isica, Universidad del Valle,\\
Ciudad Universitaria Mel\'endez, 760032, Santiago de Cali, Colombia\\
$^*$E-mail: alejandro.guarnizo@correounivalle.edu.co}

\begin{abstract}
We consider a model based on $p-$form kinetic Lagrangians in the context of dark energy.  The Lagrangian of the model is built with kinetic terms of the field strength for each $p$-form coupled to a scalar field $\phi$ through a kinetic function. We assume that this scalar field is responsible for the present accelerated expansion of the Universe. Since we are interested in cosmological applications, we specialize the analysis to a 4-dimensional case, using an anisotropic space-time. By studying the dynamical equations, we investigate the evolution of the dark energy density parameter, the effective equation of state and the shear induced by the anisotropic configuration. 
 
\end{abstract}


\bodymatter


\section{Motivation}

The predictions coming from the inflationary paradigm \cite{Guth:1982,Starobinsky1982} had been successfully confirmed with measures of the fluctuations in temperature of the Cosmic Microwave Background (CMB),  and probes in the Large Scale Structure (LSS) of the Universe, with  a significantly increase of precision during the last decades.  In its simple form, based on a single scalar field (the \textit{inflaton}) with a slow-roll potential, inflation predicts a statistical Gaussian distribution function and an isotropic power spectrum. However, some anomalies present in current data, need models beyond the standard slow-roll description in order to be fully addressed. These anomalies are related with statistical anisotropies and  signals of parity violation \cite{Ade:2015hxq}. \\

One simple attempt relies in the inclusion of vector fields (or $1-$forms), due to the intrinsic preferred directions they dictate. Models which couples a Maxwell kinetic term and a scalar field as $f(\phi) F^{\mu\nu} F_{\mu\nu}$ with $F_{\mu\nu}$ the field strenght of a vector field $A_{\mu}$, 
 had been studied in the context of inflation \cite{Watanabe:2009ct}, as well as modifications like $\phi F_{\mu\nu}\tilde{F}^{\mu\nu}$ with $\tilde{F}$ the dual of $F$ \cite{Dimopoulos:2012av}.  With the same spirit, the anisotropic spectrum of models including terms as $H_{\mu\nu\sigma}H^{\mu\nu\sigma}$ being $H_{\mu\nu\sigma}$ the corresponding field strength of a $2-$form field $B_{\mu\nu}$, had been considered \cite{Ohashi:2013qba}.\\

Besides the applications to cosmic inflation, general $p-$forms had potential interest to explain the current acceleration of the Universe. In particular, anisotropic dark energy coming from a quintessence field $\phi$ coupled to a vector field were studied in Ref. \citenum{Thorsrud:2012mu}. A similar analysis was carried out in Ref. \citenum{Almeida:2019iqp}, but this time using the field strength of a $2-$form coupled to the scalar $\phi$. In both references, possible scenarios where dark energy domination era is plausible after radiation and matter epochs, were found. We can go further the standard approach of Maxwell-like terms of the $p-$forms and allow for couplings between them, as in Ref. \citenum{Almeida:2018fwe},  where this construction was made.  The aim of this short paper is to begin studying the cosmological consequences of coupled $p-$forms. We will focus in the case of a 4-dimensional space-time and will consider the effect of a combination of  a $1-$form and a $2-$form fields coupled to a kinetic function of the quintessence field.\footnote{In $4$ dimensions there is a non-vanishing coupling term between a $1-$form and a $2-$form, which can support anisotropic inflation \cite{Almeida:2019xzt}. We leave the study of these term in the context of dark energy for a forthcoming publication.}

\section{ $p-$form-scalar model}\label{sec:pform}
We will consider the standard Lagrangian for a scalar field composed by its kinetic term and a potential $V(\phi)$ as:
\begin{equation}\label{eq:Lph}
{\cal L}_{\phi}  = \frac{1}{2}\partial_{\mu}\phi \partial^{\mu}\phi +  V(\phi).
\end{equation}
For the $p-$form sector we start with basic definitions. Given a $p$-form $A_{p \, \mu_1,\mu_2\cdots \mu_p}$, its dynamics is introduced by the field strength 
$F_{p \, \mu_1\mu_2\cdots \mu_{p+1}} \equiv  \partial_{[\mu_1} A_{p \, \mu_2 \mu_3\cdots \mu_{p+1}]}
$. In this simple case, the Lagrangian that we are going to construct will be built out of the appropriate combinations of the field strengths of the $p-$forms, coupled to the scalar field $\phi$ through arbitrary functions $f(\phi)$.  In four dimensions, only two terms remain \cite{Almeida:2018fwe}, thus the Lagrangian simply reads\footnote{The coupling functions $f_{i}(\phi)$ for each $p-$form are in general different. Here we assume them equal just for simplicity.}
\begin{equation}\label{eq:Lred}
{\cal L}_{p}(\phi, A_p) = f^2(\phi)\left( \frac{1}{4}F_{1\, \mu_1 \mu_2} F_{1}^{\mu_1 \mu_2}  + \frac{1}{12} F_{2\, \mu_1 \mu_2 \mu_3}F_{2}^{\mu_1 \mu_2 \mu_3} \right).
\end{equation}
Assuming standard gravity,  the action of our model can be written as
\begin{equation} \label{eq:LT}
S_p = \int \mr{d}^4 x \sqrt{-g}\left[ \frac{M_{\rm{p}}^2}{2}R - {\cal L}_{\phi} - {\cal L}_{p}(\phi, A_p)\right],
\end{equation}
where $M_{\rm{p}}$ the Planck mass and $R$ the Ricci scalar.

\section{Background equations} 

The Einstein equations could be written as
\begin{equation}\label{eq:Einst}
R_{\mu\nu} - \frac{1}{2}R g_{\mu\nu} = 8 \pi G ( T_{\mu\nu}^{m} + T_{\mu\nu}^{\phi} +  T_{\mu\nu}^{p}),
\end{equation} 
where we split the energy momentum tensor, $T_{\mu\nu}$, in three parts: $T_{\mu\nu}^{\phi}$, $T_{\mu\nu}^{p}$ and $T_{\mu\nu}^{m}$, representing the contributions of the scalar field, the $p$-forms and the standard matter, respectively. 
\begin{align}\label{eq:Tap}
T_{\mu\nu}^{\phi} &= \partial_{\mu}\phi \partial_{\nu}\phi - \frac{1}{2} g_{\mu\nu} \partial_{\sigma}\phi \partial^{\sigma}\phi - g_{\mu\nu}V(\phi),\\
T_{\mu\nu}^{p} &= f^2 \left[  F_{1 \, \nu\gamma}F_{1\, \mu}^{ \, \gamma}  +  F_{2 \, \, \; \nu}^{\alpha\beta}F_{2\, \alpha\beta\mu} - g_{\mu\nu} \left(\frac{1}{4}F_1^2 + \frac{1}{12} F_2^2 \right)  \right],
\end{align} 
with the shorthand notation  $F_1^2 \equiv F_{1, \mu_1 \mu_2}F^{1, \mu_1 \mu_2}$ and $F_2^2 \equiv F_{2, \mu_1 \mu_2 \mu_3}F^{2, \mu_1 \mu_2 \mu_3}$.  For the matter sector we assume a perfect fluid contribution  $T_{\mu\nu}^{m} =$ \rm{diag} $(\rho_f,  p_f, p_f,p_f)$ with $\rho_f$ the energy density and $p_f$ the preassure. Taking into account the Lagrangian of the scalar field $\mathcal{L}_{\phi}$ given in \cref{eq:Lph}, variation w.r.t.  $\phi$ gives 
\begin{align}\label{eq:Vphi}
 \square \phi - V_{,\phi} +   2 f f_{, \phi} \left(   F_{1}^2  +  F_{2}^2 \right) = 0,
\end{align}
with $f_{, \phi} \equiv \frac{\mr{d} f (\phi)}{ \mr{d} \phi}$. In which follows, we will use the gauge freedom $A_0 = \partial^{i}A_{i}=0$, to choose the vector field along the $x$ direction $A_1=A_{1}(t)\mr{d}x$, and the 2-form along the plane $y-z$, this is $A_{2} = A_{2}(t) \mr{d}y \wedge \mr{d}z$ \cite{Ohashi:2013qba}.  Thus, due to the rotational symmetry of $A_{1 \, \mu_1}$ and $A_{2 \, \mu_1 \mu_2}$ we use a Bianchi I metric:
\begin{equation}
\mr{d}s^2 = - \mr{d}t^2 + e^{2\alpha(t)}\left[e^{- 4\sigma(t)}\mr{d}x^2 + e^{2\sigma(t)}(\mr{d}y^2 + \mr{d}z^2) \right],
\end{equation}
being $e^{\alpha} \equiv a$, with $a$ the scale factor, and $\sigma$ the spatial shear.  The equation of motion (e.o.m.) for the fields $A_{1}$ and $A_{2}$  are
\begin{equation}
\ddot{A}_{1} + \left[ 2 \frac{f'}{f}\dot{\phi} + \dot{\alpha} + 4 \dot{\sigma} \right]\dot{A}_{1} = 0, \qquad 
\ddot{A}_{2} + \left[ 2 \frac{f'}{f}\dot{\phi} -( \dot{\alpha} + 4 \dot{\sigma}) \right]\dot{A}_{2} = 0,
\end{equation}
the solutions are simply
\begin{equation}
\dot{A}_{1} = \tilde{p}_{1} f(\phi)^{-2}e ^{-\alpha - 4 \sigma}, \qquad \dot{A}_{2} = \tilde{p}_{2} f(\phi)^{-2}e ^{\alpha + 4 \sigma},
\end{equation}
with $\tilde{p}_{1} $ and $\tilde{p}_{2} $ integration constants.  If we define the energy densities of the $p-$forms as 
\begin{equation}
\rho_{1}=\frac{f^2}{2} e^{-2\alpha+4\sigma} \dot{A}_1^2\,,\qquad 
\rho_{2}=\frac{f^2}{2} e^{-4\alpha-4\sigma} \dot{A}_2^2\,,
\end{equation}
the Friedmann equations, coming from \cref{eq:Einst}, and the e.o.m for the scalar field could be written as
\begin{align}
 \dot{\alpha}^2 
& = \dot{\sigma}^2 + \frac{1}{3M_{\rm pl}^2} \left[ \rho_m +  \rho_r+ \frac{1}{2} \dot{\phi}^2+V(\phi)+\rho_{1}+\rho_{2} \right]\,, \label{eq:Hmod4}\\
 \ddot{\alpha} 
& =-3\dot{\alpha}^2 + \frac{1}{M_{\rm pl}^2} \left[  \frac{\rho_m}{2}+  \frac{\rho_r}{3}  + V(\phi)+\frac{1}{3} \rho_{1}+\frac{2}{3} \rho_{2} \right]\,,\label{eq:Hmod5}\\
 \ddot{\sigma}
& =-3 \dot{\alpha}\dot{\sigma} +\frac{1}{M_{\rm pl}^2} \left[\frac{2}{3} \rho_{1}-\frac{2}{3} \rho_{2} \right]\,, \label{eq:Hmod6}\\
\ddot{\phi} & = -3 \dot{\alpha} \dot{\phi}-V_{,\phi}
+2\frac{f_{,\phi}}{f}\left( \rho_{1}+\rho_{2}\right),
\end{align}
where we take into account contributions of non-relativistic matter and radiation.

\section{Cosmological dynamics}
Let us introduce the  following dimensionless quantities  
\begin{equation}\label{eq:Params1}
\Sigma \equiv \frac{\dot{\sigma}}{\dot{\alpha}}, \quad X \equiv \frac{\dot{\phi}}{\sqrt{6}M_{\mr{pl}}H}, \quad Y \equiv \frac{\sqrt{V}}{\sqrt{3}M_{\mr{pl}}H}, \quad \Omega_{m}  \equiv \frac{\rho_{m}}{3  M_{\mr{pl}}^2 H^2},
\end{equation}
\begin{equation}\label{eq:Params2}
\Omega_r \equiv \frac{\rho_{r}}{3M_{\mr{pl}}^2 H^2}, \quad \Omega_1 \equiv \frac{\rho_{1}}{3M_{\mr{pl}}^2 H^2}, \quad  \Omega_2 \equiv \frac{\rho_{2}}{3M_{\mr{pl}}^2 H^2},  
\end{equation}
where $H \equiv \dot{\alpha}$. Thus, \cref{eq:Hmod4} can be written as $ \Omega_m =  1 - \Sigma^2 -\Omega_{DE}- \Omega_r$,  where $ \Omega_{DE} \equiv X^2+ Y^2+ \Omega_1 + \Omega_2$, is the dark energy density parameter . The effective equation of state (e.o.s.) is defined as $w_{\rm{eff}} \equiv -1 - \frac{2\dot{H}}{3H^2}$, where the ratio $\frac{\dot{H}}{H^2}$ can be computed from  \cref{eq:Hmod5} as
\begin{equation}
\frac{\dot{H}}{H^2}= -\frac{1}{2}\left( 3 + 3X^2 - 3Y^2 + 3\Sigma^2+ \Omega_1 - \Omega_2 + \Omega_r \right).
\end{equation}
In addition, we define the dark energy density and pressure as
\begin{align}
\rho_{\rm{DE}} & = \frac{\dot{\phi}}{2} + V(\phi) + \rho_{1}+\rho_{2} + 3 M_{\rm pl}^2 H^2 \Sigma^2,\\
p_{\rm{DE}} & = \frac{\dot{\phi}}{2} - V(\phi) + \frac{\rho_{1}}{3} - \frac{\rho_{2}}{3} + 3 M_{\rm pl}^2 H^2 \Sigma^2.
\end{align} 
The e.o.s for dark energy becomes
\begin{align}
w_{\rm{DE}} &\equiv \frac{p_{\rm{DE}}}{\rho_{\rm{DE}}} = \frac{3(X^2 - Y^2 + \Sigma^2)+\Omega_{1} - \Omega_2}{3(X^2+Y^2+\Sigma^2 + \Omega_{1} + \Omega_2)}.
\end{align}
In order to get a closed system of equations, is necessary to define explicitly the form of the potential $V(\phi)$ and the coupling function $f(\phi)$. We choose them to be of exponential type $V(\phi) \propto e^{-\frac{\lambda \phi}{M_{\rm{pl}}}}$, $f(\phi) \propto  e^{-\frac{\mu \phi}{M_{\rm{pl}}}} $, where $\lambda$ and $\mu$ are dimensionless  constants \cite{Thorsrud:2012mu,Almeida:2019iqp}. Thus, by differentiating w.r.t the number of $e-$folds $N \equiv \ln a$ each one of the variables given in \cref{eq:Params1,eq:Params2}  we find
{\small
\begin{align}
\Sigma' & =\frac{\Sigma}{2} \left( 3 X^2 -3Y^2+3 \Sigma ^2-3+  \Omega _1-  \Omega _2 + \Omega _r \right) +2 \Omega _1-2 \Omega _2, \label{eq:DynSys1}\\
X' &  = \frac{3X}{2}\left(X^2-Y^2+\Sigma^2-1 + \frac{\Omega_1}{3} - \frac{\Omega_2}{3} + \frac{\Omega_r}{3}\right) -\sqrt{6}\left( \mu(\Omega_1+\Omega_2) - \frac{\lambda Y^2}{2}\right),\label{eq:DynSys2} \\
Y' & =\frac{Y}{2}\left( 3X^2 - 3Y^2 + 3 \Sigma^2 + 3 + \Omega_1 - \Omega_2 + \Omega_r - \sqrt{6}\lambda X\right),  \label{eq:DynSys3} \\
\Omega_1' & =\Omega _1 \left( 3X^2 - 3Y^2 +3 \Sigma ^2+4 \Sigma-1 +2 \sqrt{6} \mu  X+\Omega _1-\Omega _2+ \Omega _r \right), \label{eq:DynSys4} \\
\Omega_2' & =\Omega _2 \left(3 X^2-3 Y^2+3 \Sigma ^2+4 \Sigma+1 +2\sqrt{6} \mu  X+\Omega _1-\Omega _2+\Omega _r \right),\label{eq:DynSys5} \\
\Omega_r' & = \Omega _r \left(3 X^2-3 Y^2+3 \Sigma ^2-1+\Omega _1-\Omega _2+\Omega _r \right).\label{eq:DynSys6}
\end{align}
}
\normalsize
\begin{figure}[t]
\centering
\begin{minipage}{0.49\textwidth}
\centering
\subfloat[]
{\label{fig:1a}\includegraphics[width=\linewidth]{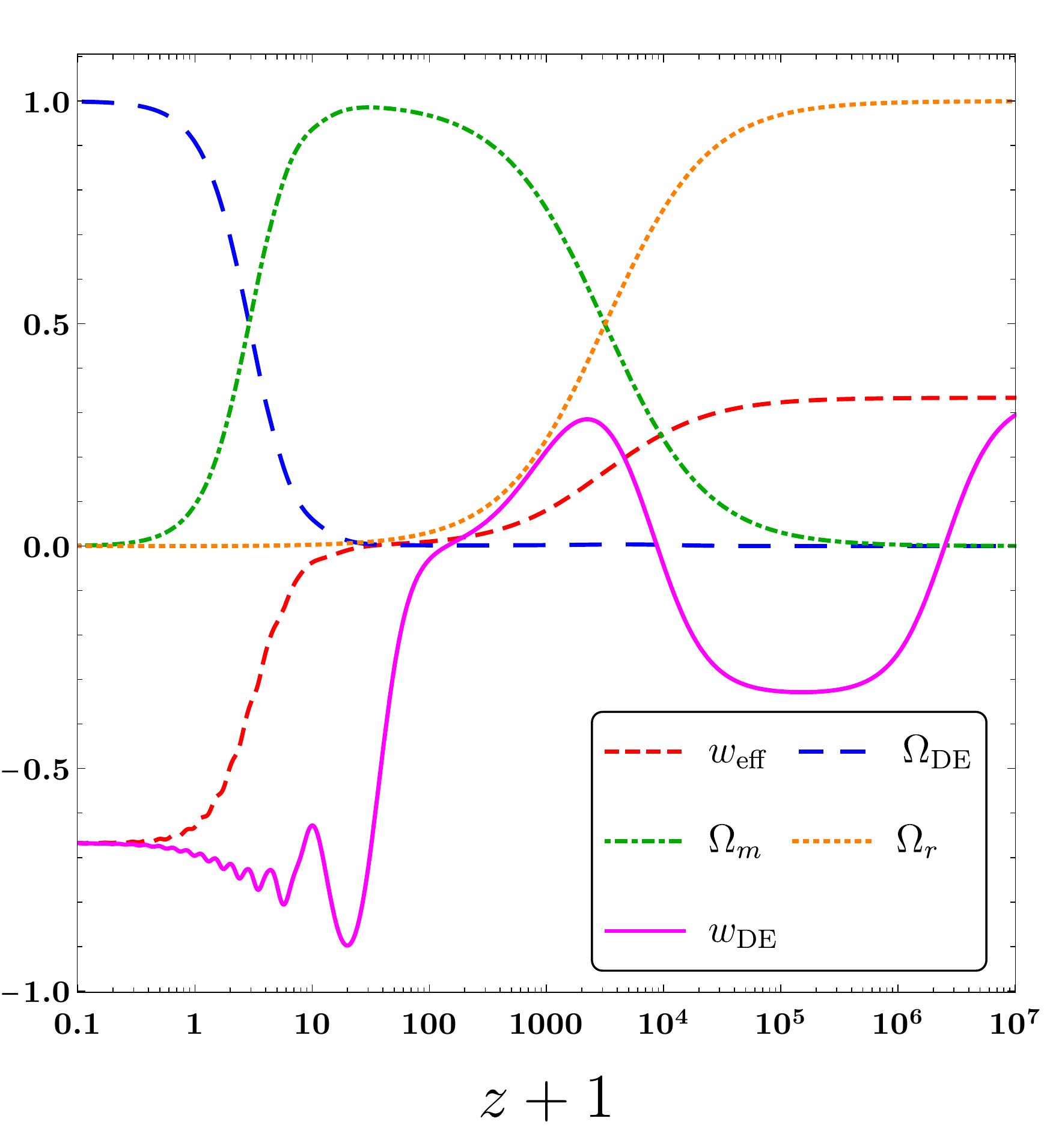}}
\end{minipage}
\begin{minipage}{0.49\textwidth}
\centering
\subfloat[]
{\label{fig:1b}\includegraphics[width=\linewidth]{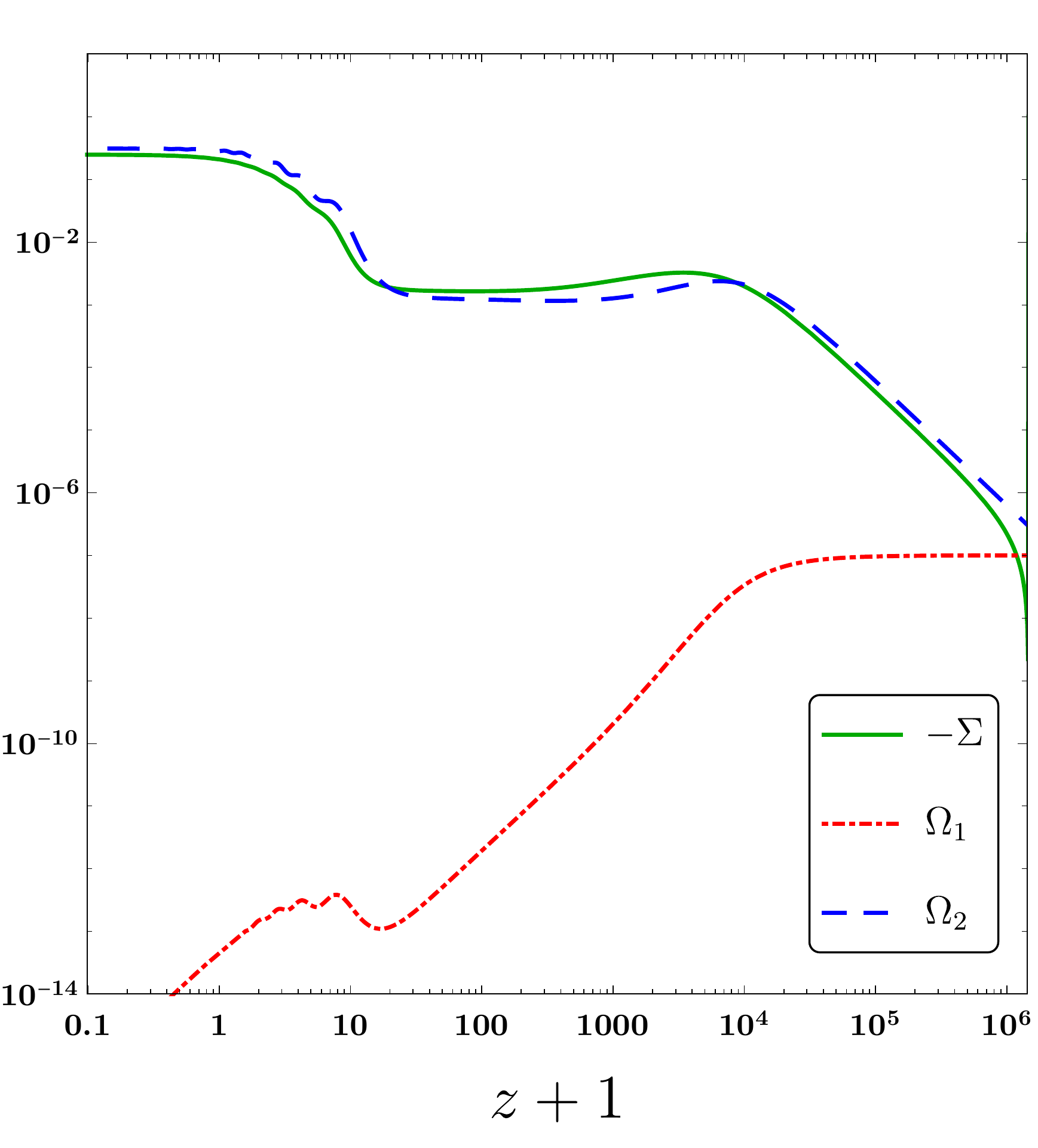}}
\end{minipage}
\caption{\label{fig:fig1} (a) Evolution of $\Omega_{\rm DE}$, $\Omega_r$, $\Omega_m$, 
$w_{\rm DE}$, and $w_{\rm eff}$ versus $z+1$ for $\lambda=\mu=10$, with the initial conditions $X=10^{-13}$, $Y=10^{-14}$, $\Sigma=0$, $\Omega_1=10^{-7}$, $\Omega_2=10^{-10}$, and $\Omega_r=0.99996$ at the redshift $z=7.9 \times 10^7$. (b) Evolution of  $-\Sigma$,  $\Omega_1$ and $\Omega_2$  versus $z+1$ 
for $\lambda=\mu=10$ with 
the same set of initial conditions as figure (a). }
\end{figure}
\noindent Instead of the standard analysis of critical points and stability for the previous autonomous system, we decide the make numerical integrations of the equations to obtain a general behavior of the solutions. A complete analysis of this system goes beyond this short paper and is left for a forthcoming work. In \cref{fig:fig1} we shown the numerical integration of the set of  \cref{eq:DynSys1,eq:DynSys2,eq:DynSys3,eq:DynSys4,eq:DynSys5,eq:DynSys6} where the couplings constants were fixed to be $\lambda = \mu=10$. Typically, we search for a transition from radiation dominance, to matter dominance, and finllay reach an epoch dominated by dark energy today. In terms of effective equation of state $w_{\rm{eff}}$ those transitions are of  the type $w_{\rm{eff}}\sim \frac{1}{2} \longrightarrow w_{\rm{eff}}\sim 0  \longrightarrow w_{\rm{eff}} \leq - \frac{1}{3} $, as we can see in figure \cref{fig:1a}, where the sequence is observed in terms of the density parameters. For large redshifts the radiation dominates and start to decrease at an approximate redshift of $z \sim 3000$ to a matter dominated epoch ($ \Omega_{m} \sim 1$), as expected. The evolution of the shear $\Sigma$ and the density parameters  $\Omega_1 $ and $\Omega_2$ are shown in \cref{fig:1b}. The contribution of the $2-$form always dominate over the $1-$form, except for the initial condition at high redshift ($\sim 10^{7}$); the shear increases until a constant value in the present time. In contrast with the case presented in Ref. \citenum{Almeida:2019iqp}, where only the $2-$form is considered, $w_{\rm{eff}}$ and $w_{\rm{DE}}$ we will not reach the asymptotic value $\sim - 1$. Nevertheless, we check numerically by evolving $w_{\rm{DE}}$ with different (larger) values of the coupling $\mu$, that this value is realized. \\

\noindent As we anticipated, a complete analysis including the stability of fixed points of the autonomous system, will be presented in a future work, where we also want to elucidate the effect of a non-vanishing coupling between the $1-$ and $2-$ form fields.

\section*{Acknowledgments}
This work was supported by COLCIENCIAS Grant No. 110656933958 RC 0384-2013 and by COLCIENCIAS grant 110278258747 RC-774-2017 (DAAD-Procol program).  AG also acknowledges financial support from the MG15 organizing committee and from the ESA travel funds.


\end{document}